\documentclass[prl,twocolumn,showpacs,superscriptaddress]{revtex4}

\usepackage{amssymb,amsfonts,amsmath} 
\usepackage{graphicx}
\usepackage{bm}
\usepackage{color}
\usepackage{hyperref}

\newcommand{\cprl}[3]{Phys.~Rev.~Lett.~{\bf #1}, #2 (#3)}

\begin{document} \title{Luttinger liquid universality in the time evolution after an interaction quench}

\author{C.\ Karrasch} 
\affiliation{Department of Physics, University of California, Berkeley, California 95720, USA}

\author{J.\ Rentrop}  
\affiliation{Institut f{\"u}r Theorie der Statistischen Physik, RWTH Aachen University 
and JARA---Fundamentals of Future Information
Technology, 52056 Aachen, Germany}

\author{D.\ Schuricht}  
\affiliation{Institut f{\"u}r Theorie der Statistischen Physik, RWTH Aachen University and JARA---Fundamentals 
of Future Information
Technology, 52056 Aachen, Germany}

\author{V.\ Meden} 
\affiliation{Institut f{\"u}r Theorie der Statistischen Physik, RWTH Aachen University and JARA---Fundamentals 
of Future Information
Technology, 52056 Aachen, Germany}

\begin{abstract} 

We provide strong evidence that the relaxation dynamics of one-dimensional, metallic 
Fermi systems resulting out of an abrupt amplitude change of the two-particle 
interaction has aspects which are universal in the Luttinger liquid sense: The 
leading long-time behavior of certain observables is described by universal 
functions of the equilibrium Luttinger liquid parameter and the renormalized velocity. 
We analytically derive those functions  for the Tomonaga-Luttinger model and verify our 
hypothesis of universality by considering spinless lattice fermions within the 
framework of the density matrix renormalization group.        
  
\end{abstract}

\date{\today} 
\maketitle

The equilibrium low-energy physics of a large class of one-dimensional (1d), correlated, 
metallic Fermi systems is described by the Luttinger liquid (LL) 
phenomenology \cite{Giamarchi03,Schoenhammer05}. The Tomonaga-Luttinger (TL) model is the 
effective low-energy fixed point model of the LL universality class and thus 
plays the same role as the free Fermi gas in Fermi liquid theory.  This universality
relies on the renormalization group (RG) irrelevance of contributions such as the 
momentum dependence of the two-particle interaction \cite{Meden99} or the curvature of the 
single-particle dispersion \cite{Imambekov11} which are present 
in microscopic models but ignored in the TL model. 
For a model falling into the LL universality class it 
is not necessary to explicitely compute thermodynamic observables
and correlation functions if 
one is interested in the low-energy limit. One only needs to determine two 
numbers -- the LL parameter $K$ and the renormalized velocity $v$ of the 
excitations -- which fully characterize the low-energy physics of a spinless 
LL (on which we focus). Those depend on the band structure and filling 
as well as the amplitude and range of the two-particle interaction 
of the microscopic model at hand and can be extracted from the ground-state 
energy \cite{Haldane80} or `simple' response functions \cite{llparam}. 
Thereafter, correlation functions at long length scales or thermodynamic quantities 
at low energies can be obtained by plugging in $K$ and $v$ into analytic 
expressions derived within the exactly solvable 
TL model \cite{Giamarchi03,Schoenhammer05}. 

The recent progress in experimentally controlling isolated many-body 
states, in particular in cold atomic gases \cite{Bloch08}, 
led to numerous theoretical studies on the dynamics of closed 
quantum systems resulting out of an abrupt change of the amplitude  $U$ 
of the two-particle interaction \cite{Polkovnikov11}. 
One assumes that the system is prepared in a canonical thermal state or, 
at temperature $T=0$ on which we focus, the ground state of an initial 
Hamiltonian $H_{\rm i}$ with a given $U_{\rm i}$; often $U_{\rm i}=0$ is 
considered. At time $t=0$ the interaction is quenched to $U_{\rm f}$ and 
the time evolution is performed with the final Hamiltonian $H_{\rm f}$. 
Fundamental questions discussed are \cite{Polkovnikov11}: (a) Do some observables become 
stationary at large times? (b) How can they be classified (locality)? (c) Is it 
possible to compute their steady-state expectation values using an 
appropriate density matrix $\rho_{\rm st}$? It was conjectured that in models 
with many integrals of motion, e.g.~those which are solvable by 
Bethe ansatz (`integrable') \cite{Faribault09,Mossel10,Gritsev10}, 
$\rho_{\rm st}$ is not of thermal but rather 
of generalized (`Gibbs') canonical form \cite{Rigol07}. This 
has been confirmed for models which can be mapped to effective noninteracting 
ones \cite{Rigol07,Calabrese07,Eckstein08,Kollar08,Rossini09,Calabrese11,Barthel08}, in particular 
the TL model \cite{Cazalilla06} and a variant of the latter \cite{Kennes10}, 
but a general prove is lacking. For generic models one generally 
expects $\rho_{\rm st}$ to be thermal. 

Here we address questions about the \textit{relaxation dynamics towards 
the steady state}. Considering 
several models which in equilibrium fall into the LL class we ask: (A) 
{\em Can the time evolution be characterized as universal in the LL sense?} 
As model-dependent high energy processes 
matter at short to intermediate times one can only expect to find LL universality 
on \textit{large time scales} -- but even then it is not obvious whether the notion of RG 
irrelevance can be transferred from equilibrium to the nonequilibrium 
dynamics \cite{Mitra11,Rentrop12}. 
(B) {\em If LL universality is found, does it hold independently of 
the number of integrals of motion and thus independently of the expected nature of 
the steady state} (generalized canonical versus thermal)? 
To answer those questions we proceed in steps. We compute the time evolution of 
the $Z${\em -factor} jump of the momentum 
distribution function $n(k,t)$ at the Fermi momentum $k_{\rm F}$ and the 
{\em kinetic energy per length} $e_{\rm kin}(t)$ for 
the spinless TL model with arbitrary momentum dependent 
interactions $g_{2/4}(k)$ using bosonization \cite{Giamarchi03,Schoenhammer05}. $e_{\rm kin}(t)$ is defined as the expectation 
value of the initial Hamiltonian $H_{\rm i}$ and thus describes how excitations die out. The 
quench is performed out of the noninteracting ground state.  
We analytically show that for any continous $g_{2/4}(k)$ the long-time 
dynamics are given by \cite{footnote0} 
\vspace{-0.3cm}
\begin{subequations}\label{asymptotics}
\begin{align}
Z(t) & \sim t^{-\gamma_{\rm st}(K)}~, \label{asymptotics1} \\
\left|de_{\rm kin}(t)/dt
\right| &\sim \epsilon(K,v)t^{-3} ~. \label{asymptotics2}
\end{align}
\end{subequations}
The decay of $Z(t)$ is governed by a universal exponent $\gamma_{\rm st}(K)$ that depends on the equilibrium LL parameter $K$ only; $e_{\rm kin}(t)$ features an asymptotic power law with an interaction-independent exponent but universal prefactor $\epsilon(K,v)$ determined by $K$ as well as by the renormalized velocity $v$. In equilibrium, this is the characteristic $K$- and $v$-dependence of correlation functions or of thermodynamic quantities, which \textit{a posteriori} motivates to consider $Z(t)$ and $e_{\rm kin}(t)$ as representative examples. The notion of LL universality can now be defined in analogy to equilibrium: 
\textit{The quench dynamics is universal if 
Eqs.~(\ref{asymptotics}) describe the long-time relaxation for any model falling into the 
equilibrium LL universality class} if the corresponding values for $K$ and 
$v$ are plugged in. To investigate this we compute $Z(t)$ for a 1d lattice of 
spinless fermions with nearest-neighbor hopping 
and interaction $\Delta$ \cite{Manmana07} as well as an extension of the latter 
including a next-to-nearest-neighbor 
interaction $\Delta_2$. We use the numerical time-dependent density-matrix 
RG (DMRG) \cite{Schollwoeck11,white,tdmrg}. The model with $\Delta_2=0$ has 
many conserved quantities, is Bethe ansatz integrable, and thus $K$ as well 
as $v$ are known analytically \cite{Haldane80,Qin97}. The $\Delta$-$\Delta_2$--model, however, 
is believed to be not exactly solvable. For $\Delta_2>0$ we extract $K$ and $v$ 
from equilibrium quantities (e.g.~the small momentum density response 
function) using DMRG \cite{footnote2,llparam,LLpaper}. Our data for the $Z$-factor 
agrees with Eq.~(\ref{asymptotics1}) for any interaction strength, filling factor, 
and irrespective of the integrability of the model. The results for $e_{\rm kin}(t)$ 
are consistent with Eq.~(\ref{asymptotics2}), but on the time scales accessible 
by DMRG the asymptotic behavior is still masked 
by oscillatory terms of higher order in $t^{-1}$. To unambiguously determine 
the prefactor of the $t^{-3}$-decay of the energy we resort to 
a numerical trick. Instead of performing the time evolution with $\exp{(-i H_{\rm f} t)}$ 
we apply the imaginary time analogue $\exp{(- H_{\rm f} \tau)}$.
In this case the {\em total} energy per length $e(\tau)$ -- which is no 
longer conserved -- is the natural observable. For the TL model we show 
that the asymptotics is completely analogous to Eq.~(\ref{asymptotics2}) 
with $t \to \tau$ and $\epsilon(K,v)$ replaced by a different function 
$\epsilon_{\rm it}(K,v)$. For the lattice model, the $\tau^{-3}$-decay 
manifests over several orders of magnitude, and the prefactor agrees 
with the TL prediction.

This altogether provides strong evidence that questions (A) and (B) 
can be answered by `yes'. We conjecture that the universality of the quench 
dynamics also holds for other models falling into the equilibrium LL class.           

\begin{figure}[t]
\includegraphics[width=0.95\linewidth]{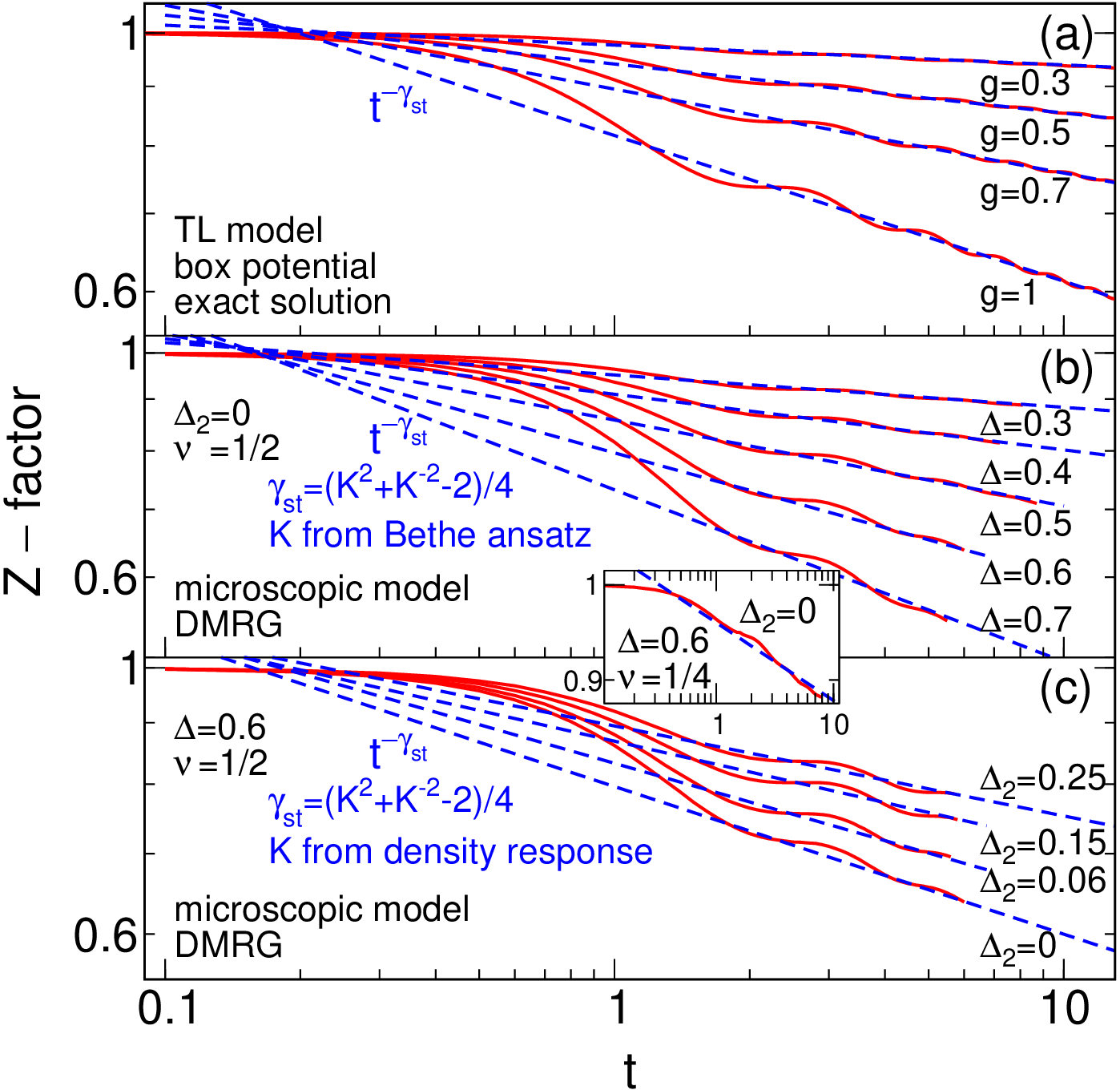}\vspace*{-0.2cm}
\caption{(Color online) Time evolution of the $Z$-factor out of the noninteracting ground state 
of a 1d metallic Fermi system after switching on two-particle terms at time $t=0$. 
Dashed lines show the \textit{universal asymptotic power law} $t^{-\gamma_{\rm st}(K)}$ 
with an exponent determined by the equilibrium LL parameter $K$. (a) TL
model. The plots displays box-like two-particle interactions $g(k)$ of strength $g=g(0)$; 
the asymptotics are universal for any $g(k)$. Time is given in units of $(v_{\rm F}k_{\rm c})^{-1}$. 
(b, c, Inset) Spinless lattice fermions 
of Eq.~(\ref{Hlatticemodel}) at filling $\nu$ featuring nearest ($\Delta$) and next-nearest ($\Delta_2$) 
neighbor interactions. }
\label{fig:zfac}
\end{figure}

\begin{figure*}[t]
\includegraphics[width=0.46\linewidth,clip]{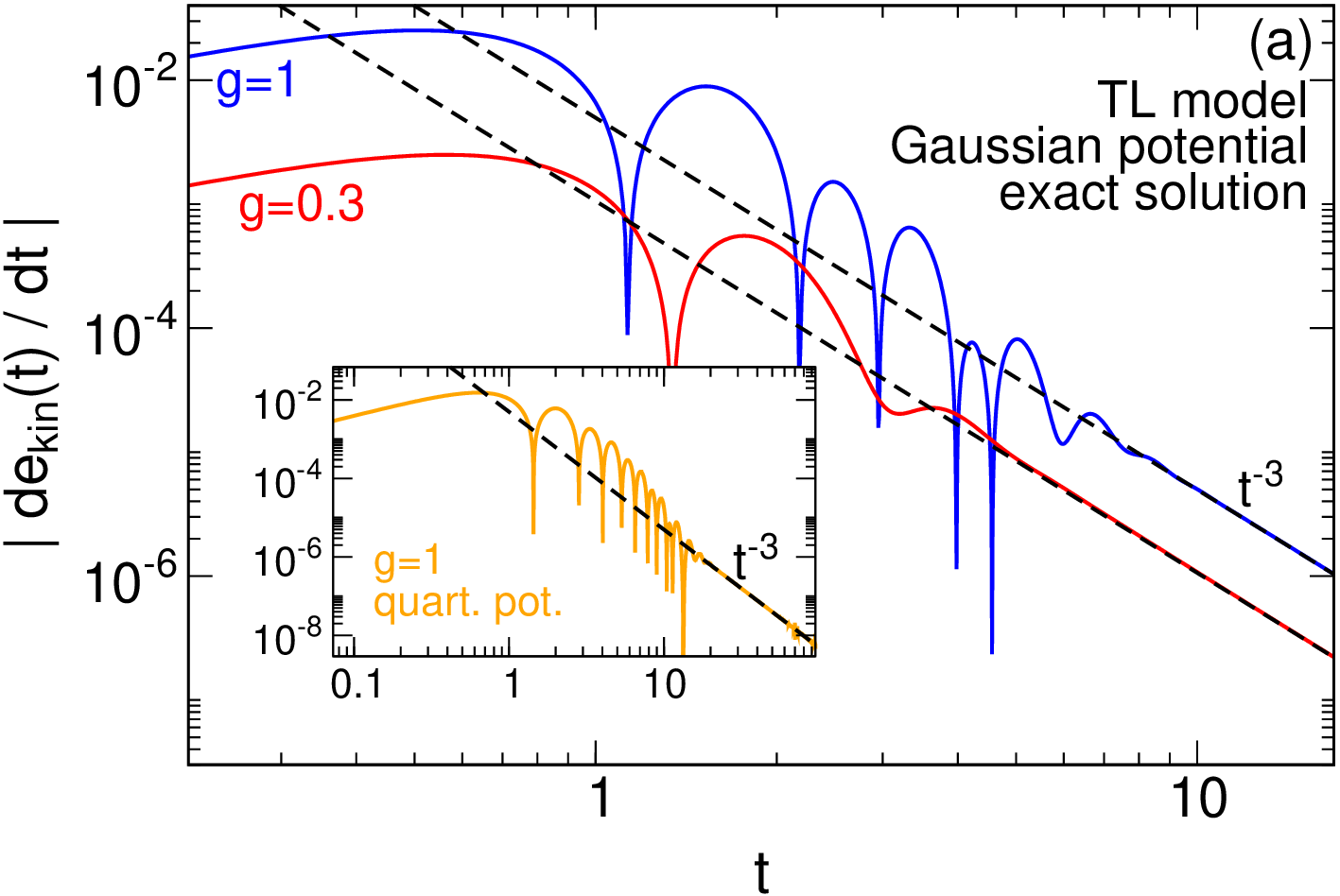}\hspace*{0.04\linewidth}
\includegraphics[width=0.46\linewidth,clip]{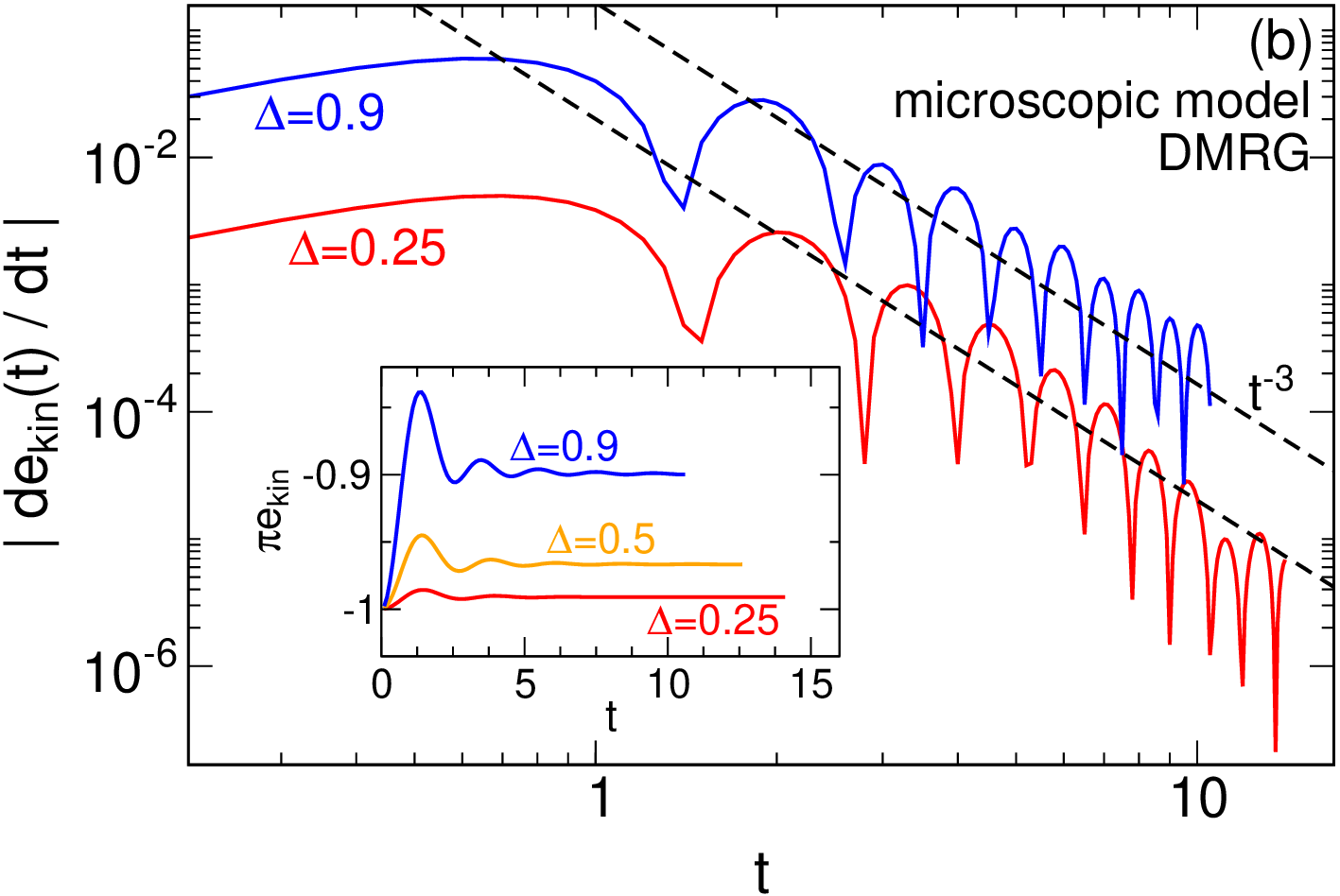}
\caption{(Color online) Time evolution of the kinetic energy per length $de_{\rm kin}/dt$. 
(a) TL model for different (Gaussian and quartic) two-particle 
potentials $g(k)$. For any continous $g(k)$, $de_{\rm kin}/dt$ 
asymptotically falls off as $\epsilon(K,v)/t^3$ with a universal 
prefactor. $e_{\rm kin}$ and $t$ are given in units of $v_{\rm F}k_c^2$ and
$(v_{\rm F}k_{\rm c})^{-1}$, respectively. (b) Spinless lattice fermions. Solid lines show DMRG data; dashed lines display $t^{-3}$ power laws where the ratio of prefactors is chosen according to the TL prediction. Inset: DMRG data before taking the $t$-derivative.}
\label{fig:ekin}
\end{figure*}

\textit{The TL model} --- After bosonizing  \cite{Giamarchi03,Schoenhammer05} 
the density of left and right moving fermions with 
a linear dispersion the Hamiltonian of the TL model is quadratic in operators 
$b_n^{(\dag)}$ which obey bosonic commutation relations:
\begin{eqnarray} 
 H  = \sum_{n >0 } &&  \left[ k_n \left( v_{\rm F} + \frac{g_4(k_n)}{2 \pi} \right) 
\left( b_n^\dag b_n^{} + b_{-n}^\dag b_{-n}^{} \right) \right. \nonumber \\
&& \left. +   k_n   \frac{g_2(k_n)}{2 \pi} 
\left( b_n^\dag b_{-n}^\dag + b_{-n}^{} b_{n}^{} \right)  \right] ~,
\label{HTL}
\end{eqnarray}
where $k_n=2 \pi n / L$, $n \in {\mathbb Z}$, $L$ denotes the chain length, and $v_{\rm F}$ 
is the Fermi velocity. The two coupling functions (potentials) $g_{2/4}$ determine the 
strength of the scattering of fermions 
on different branches ($g_2$) and the same branch ($g_4$). Usually the $k$-dependence of 
$g_{2/4}$ is neglected and integrals are regularized in the ultraviolet through an ad hoc 
procedure \cite{Giamarchi03,Schoenhammer05}. As the momentum dependence is RG irrelevant 
this is justified in equilibrium if all energy scales are sent to zero \cite{Meden99}. 
For the quench dynamics -- even at asymptotic times -- it is, however, not clear if the same 
reasoning holds and we thus 
keep the full $k$-dependence and consider coupling {\em functions.} In fact, it was 
recently shown 
that the momentum dependence indeed affects the long-time dynamics of certain 
observables \cite{Rentrop12}. For the system 
to be a LL in equilibrium we require that $0 < g_{2/4}(0) < \infty $ 
(repulsive interactions) and that 
$g_{2/4}(k)$ decay on a scale $k_{\rm c}$. The Hamiltonian of Eq.~(\ref{HTL}) 
can be diagonalized to $H  = \sum_{n \neq 0} \omega(k_n) \, \alpha_n^\dag \alpha_n^{}  
+ E_{\rm gs}$ by introducing new modes $\alpha_n = c(k_n) b_n + s(k_n) b^\dag_{-n}$ with
\begin{eqnarray} 
&& s^2(k) = \frac{1}{2} \left[ \frac{1+ \hat g_4(k)}{W(k)} -1  \right]  = c^2(k) -1   
~,\label{manydefs}  \\  && 
\omega(k) = v_{\rm F}  |k| \, W(k) = v_{\rm F}  |k|\sqrt{(1+ \hat g_4(k))^2 - \hat g_2^2(k)}  \nonumber ~, 
\end{eqnarray}    
where $\hat g_{2/4}= g_{2/4}/(2 \pi v_{\rm F})$, and $E_{\rm gs}$ denoting the ground state energy. The LL parameter and the renormalized velocity read
\begin{eqnarray} 
K= \sqrt{\frac{1+\hat g_4(0) -\hat g_2(0)}{1+\hat g_4(0) +\hat g_2(0)}}~ , \; \; \; \; v=v_{\rm F} W(0)~ .
\label{Kv}
\end{eqnarray}
As our initial state we take the noninteracting ground state $\left| E_{\rm gs}^{0} \right\rangle$ 
which is given by the vacuum $\left| \mbox{vac}(b) \right>$ with respect to 
the $b_n$. Expectation values of the time-evolved state $  \left|  \Psi(t) \right\rangle 
= \exp(-i H t) \left|  E_{\rm gs}^{0} \right\rangle$ can be computed straightforwardly 
using the simple time dependence of the eigenmode operators $\alpha_n^{(\dag)}$ 
and their linear dependence on the $b_n^{(\dag)}$ \cite{Rentrop12}.  
 
After bosonizing the fermionic field operator \cite{Giamarchi03,Schoenhammer05} 
the $Z$-factor $Z(t)=\lim_{k\nearrow k_{\rm F}}n(k,t)-
\lim_{k\searrow k_{\rm F}}n(k,t)$ is easily obtained (taking $L \to \infty$) 
\cite{Cazalilla06,Rentrop12,Dora11}:
\begin{eqnarray}
Z(t) = \exp{ \left\{ - \int_0^\infty  \!\!\! dk \, \frac{4 s^2(k) c^2(k)}{k} 
\left(  1- \cos{\left[2 \omega(k) t\right]} \right)  \right\} } .
\nonumber
\end{eqnarray}
Independent of the form of $g_{2/4}(k)$ (even for potentials with a discontinous jump to zero 
at $k_{\rm c}$) the large-time behavior is given by Eq.~(\ref{asymptotics1}) with 
$\gamma_{\rm st} = (K^2 + K^{-2}-2)/4$; it manifests on the (nonuniversal) scale 
$(v_{\rm F}k_{\rm c})^{-1}$. Figure \ref{fig:zfac}(a) shows $Z(t)$ obtained by 
numerically performing the 
integral for a simple box shaped potential $\hat g_2(k) = \hat g_4(k) = g  
\Theta(k_{\rm c} - |k|)/2$ of varying amplitude $g$. The asymptotic power-law is modulated by 
oscillations which decay faster than $t^{- \gamma_{\rm st}}$.

The kinetic energy per length $e_{\rm kin}(t)$ reads ($L \to \infty$)  
\begin{eqnarray}
e_{\rm kin}(t)  =  \frac{ v_{\rm F}}{2 \pi} \!\! \int_0^\infty \!\!\!\! dk k 4 s^2(k) c^2(k)  
\left\{ 1- \cos{[2 \omega(k) t]} 
\right\} .
\label{EkintTD}
\end{eqnarray} 
The steady-state value is obtained by dropping the oscillatory term which averages 
out for $t\to\infty$. For {\em continuous} 
coupling functions $g_{2/4}(k)$ of range $k_{\rm c}$ asymptotic analysis yields 
Eq.~(\ref{asymptotics2}) as the leading term in the long-time limit;\cite{footnote4} the coefficient 
is given by $\epsilon(K,v)=\gamma_{\rm st}(K) v_{\rm F}/(4 \pi v^2)$. 
Figure \ref{fig:ekin}(a) shows the derivative of 
$e_{\rm kin}$ for $\hat g_2(k) = \hat g_4(k) = g(k)$, a Gaussian potential 
$g(k)=g \exp(-[k/k_{\rm c}]^2/2) /2$ as well as a quartic potential 
$g(k) = g/(1+[k/k_{\rm c}]^4)/2$ and varying interaction strengths. 
As either $g(0)$ or the lowest nonvanishing Taylor expansion order of $g(k)-g(0)$ increases, 
the amplitude of an oscillatory term which decays faster than the leading one becomes stronger. 
The (nonuniversal) scale on which the asymptotic $t^{-3}$-behavior dominates thus heavily 
depends on the strength and 
type of potential at hand [compare the inset and the main part
of Figure \ref{fig:ekin}(a)]. 

\textit{Microscopic lattice model} --- As a next step we provide strong evidence that 
Eqs.~(\ref{asymptotics}) describe the long-time relaxation dynamics of any model which 
\textit{in equilibrium} falls into the LL universality class. To this end, we consider 
spinless lattice fermions,
\begin{eqnarray}
\mbox{} \hspace{-.3cm} H \! = \! \sum_{j} \! \left[ \frac{1}{2} c_j^{\dag} 
c_{j+1}^{\phantom{\dagger}} + \mbox{H.c.} + \Delta n_j n_{j+1} + \Delta_2 n_j n_{j+2} 
\right] ,
\label{Hlatticemodel}
\end{eqnarray}
with $n_j = c_j^\dag c_j-1/2$. We study the 
quench dynamics using an infinite-system DMRG 
algorithm \cite{Schollwoeck11,tebd}. We determine 
$|E_{\rm gs}^0\rangle$ by applying an imaginary time evolution 
$\exp(-\tau H|_{\Delta=\Delta_2=0})$ to a random initial 
matrix product state with a fixed matrix dimension $\chi$ 
until the energy has converged to typically $8-10$ relative digits. 
Operators $\exp(\sim H)$ are factorized by a second or fourth order 
Trotter decomposition. Thereafter, we compute the real time evolution 
$|\Psi(t)\rangle=\exp(-itH)|E_{\rm gs}^0\rangle$ in presence of the 
two-particle terms $\Delta$ and $\Delta_2$. $\chi$ is dynamically 
increased in order to maintain a fixed discarded weight. We carefully 
ensure that the latter is chosen small enough (and that the initial $\chi$ is 
large enough) to obtain numerically exact results.

\begin{figure}[t]
\includegraphics[width=0.95\linewidth,clip]{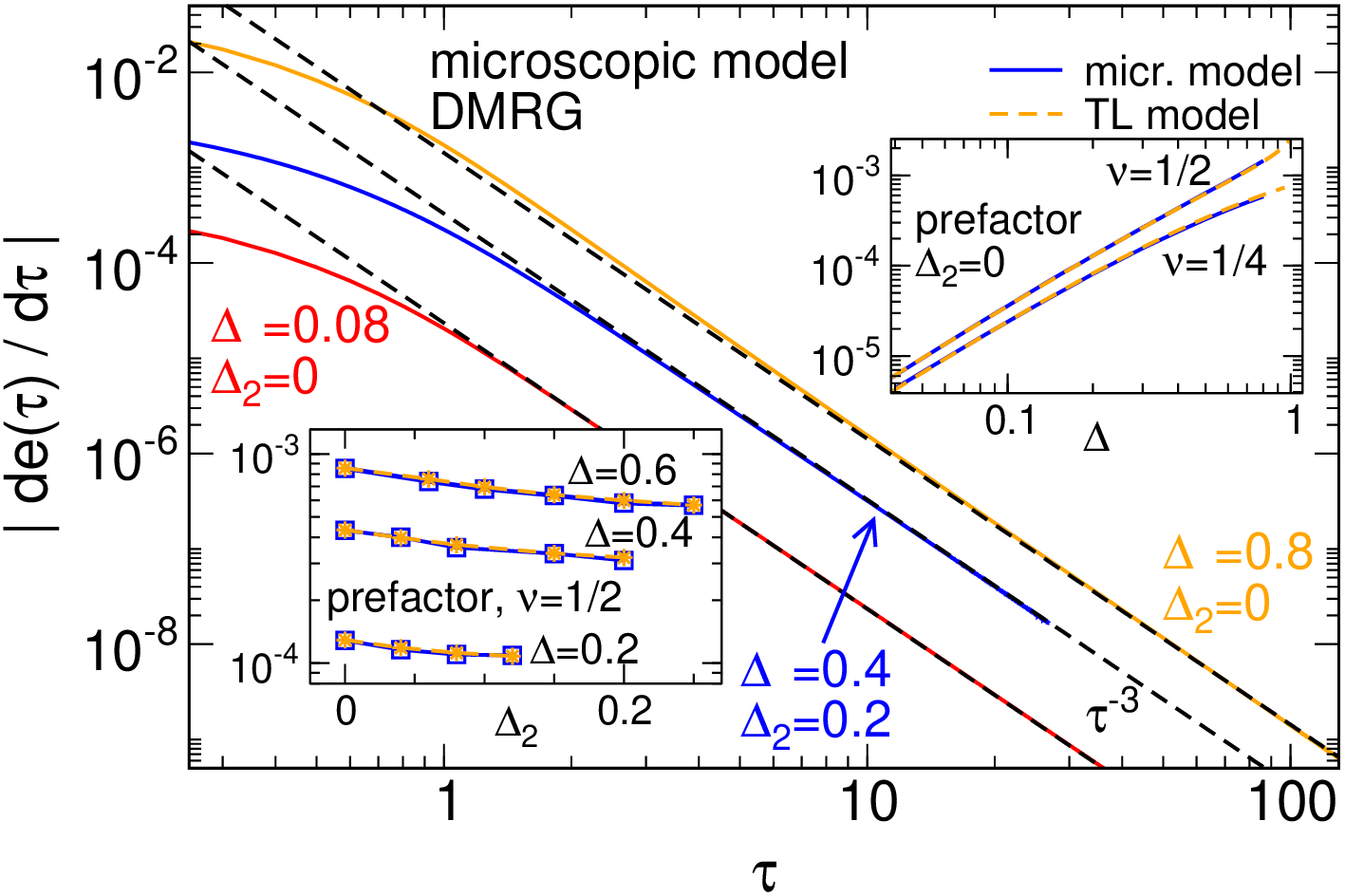}
\caption{(Color online) Imaginary time evolution $\exp(-\tau H)|E_{\rm gs}^0\rangle\stackrel{\tau\to\infty}
{\longrightarrow}|E_{\rm gs}\rangle$ towards the \textit{interacting} ground state 
$|E_{\rm gs}\rangle$. The main part shows DMRG data for the $\tau$-derivative of the total 
energy  in the lattice model. The $\tau^{-3}$-decay predicted 
by bosonization 
manifests over several orders of magnitude. The prefactor agrees with the TL formula 
$\epsilon_{\rm it}(K,v)$ for all parameters (this is illustrated in the insets). The 
imaginary-time energy dynamics are thus universal in the LL sense.}
\label{fig:imag}
\end{figure}

The time evolution of the momentum distribution function $ n(k,t) = 
\sum_j e^{ikj} \big\langle\Psi(t)\big| c_j^\dagger c_j^{\phantom{\dagger}}
\big|\Psi(t)\big\rangle$ and the corresponding $Z$-factor can be computed 
straightforwardly; the $j$-sum is 
carried out up to $\sim10000$ sites. Results for $Z(t)$  are shown in 
Figure \ref{fig:zfac}(b,c). Its long-time asymptotics indeed shows a power-law 
decay $t^{-\gamma_{\rm st}(K)}$, and the exponent agrees to the one predicted 
by the TL formula (dashed lines) if the latter is evaluated for $K$ corresponding to 
the microscopic parameters under consideration. 
We take $K$ from the Bethe ansatz ($\Delta_2=0$) \cite{Haldane80,Qin97} or the 
equilibrium density response 
($\Delta_2>0$) \cite{llparam,LLpaper}. The agreement with the TL model result 
holds for any interaction strength \cite{footnote2}, for any 
filling factor $\nu$, and irrespective of the integrability of the model. 
This strongly indicates that the asymptotic dynamics of the $Z$-factor is indeed 
universal in the LL sense.

The time derivative of the kinetic energy (the expectation value of 
$H|_{\Delta=\Delta_2=0}$) 
per length (site) is shown in Figure \ref{fig:ekin}(b). Its magnitude decays 
as $t^{-3}$, and the ratio between prefactors (where the factor 
$v_{\rm F}$ drops out) at different interaction strengths is consistent with the TL 
formula (see the dashed lines). However, oscillations have not died out completely, 
and a pure power law cannot be identified unambigously. To further support 
that this is merely because the time scales reachable in our DMRG calculation 
are too small -- remind that for the TL model the scale where $de_{\rm kin}/dt$ is 
governed by Eq.~(\ref{asymptotics2}) strongly depends on the strength and type 
of the potential in contrast to $Z(t)$ where it is always $(v_{\rm F}k_{\rm c})^{-1}$ -- 
and that the energy relaxation is indeed universal, we consider an 
imaginary time evolution, $\left| \Psi(\tau) \right\rangle 
= \exp{(- H \tau)}\left| E_{\rm gs}^{0}   \right\rangle/\left\langle E_{\rm gs}^{0} \right| 
\exp{(- 2 H \tau)}  \left|  E_{\rm gs}^{0} \right\rangle^{1/2}$. For $\tau\to\infty$, 
$\left| \Psi(\tau) \right\rangle$ approaches the ground state $ \left|  E_{\rm gs} 
\right\rangle$ of the interacting Hamiltonian. The total energy is 
the natural observable to compute in this academic scenario. Its asymptotic 
behavior within the TL model is completely 
analogous to Eq.~(\ref{asymptotics2}) 
with $t \to \tau$ and $\epsilon(K,v) \to \epsilon_{\rm it}(K,v) = 
\mbox{Li}_2([K+K^{-1}-2]/[K+K^{-1}+2])/(8 \pi v)$, 
where $ \mbox{Li}_2$ denotes the 
dilogarithm \cite{footnote3}. For the lattice model, one can easily access large imaginary times 
using DMRG, and the $\tau^{-3}$-decay manifests over several orders of magnitude. 
This is illustrated in Fig.~\ref{fig:imag}. The prefactor (shown in the insets) agrees with 
$\epsilon_{\rm it}(K,v)$ (the latter depends on $K$ and $v$ only; thus, one does not need to 
consider ratios) for all interactions and fillings. The dynamics of the total energy at 
large $\tau$ is universal.

\textit{Conclusion} --- We have obtained exact expressions for the time evolution 
of the $Z$-factor and of the kinetic energy after an interaction quench within the 
Tomonaga-Luttinger model. For any continous two-particle potential, their long-time 
asymptotes $Z\sim t^{-\gamma_{\rm st}}$, $de_{\rm kin}/dt\sim \epsilon(K,v)/t^3$ are 
universal functions of the LL parameter $K$ and the renormalized velocity $v$. 
We studied a similar scenario for spinless lattice fermions using DMRG; for large 
times, $Z(t)$ and $e_{\rm kin}(t)$ are described by the above expressions. This
provides strong evidence that the relaxation dynamics after an interaction 
quench within any model that falls into the equilibrium LL universality class 
has aspects which are universal in the LL sense.

\textit{Acknowledgments} --- We thank S.~Kehrein, D.~M.~Kennes, J.~E.~Moore, and 
K.~Sch\"onhammer for fruitful discussions.  This work was supported by the DFG via 
KA3360-1/1 (CK), the Emmy-Noether program (DS), and FOR 912 (VM).

{}


\begin{thebibliography}{}
\bibitem{footnote0} To eliminate the steady-state value of $e_{\rm kin}(t)$ we take the 
  derivative.
\bibitem{footnote2} We focus on parameter regimes for which the model is a LL in equilibrium 
  \cite{Giamarchi03,Schoenhammer05}.
\bibitem{footnote4}
This holds as long as the interaction is smaller than a
critical scale beyond which $\omega(k)$ of Eq.~(\ref{manydefs}) develops
stationary points \cite{Rentrop12}. This scale depends on the form of
the $g_{2/4}(k)$.
\bibitem{footnote3}
  This result can straightforwardly be obtained by writing 
$\left| \mbox{vac}(b) \right>= \prod_{n >0 } 
\sum_{l \geq 0} a_l^{(n)} \left| l \right>_{n} \left| l \right>_{-n}$, where 
$\left| l \right>_{m}$ is the $l$-th oscillator eigenstate with respect to the
eigenmode given by $\alpha_m^{(\dag)}$ and applying $\exp(- \tau H)$. The expansion 
coefficients are $a_l^{(n)}=(-1)^l [s(k_n)/c(k_n)]^l/c(k_n)$.
\end{thebibliography}
\end{document}